# A polarization-induced 2D hole gas in undoped gallium nitride quantum wells

Reet Chaudhuri[1]*, Samuel James Bader[2], Zhen Chen[2], David A. Muller[2,4],

Huili (Grace) Xing[1,3,4], Debdeep Jena[1,3,4].

[1]School of Electrical and Computer Engineering, Cornell University, Ithaca NY 14853

[2]School of Applied and Engineering Physics, Cornell University, Ithaca NY 14853

[3]Department of Materials Science and Engineering, Cornell University, Ithaca NY 14853

[4]Kavli Institute at Cornell for Nanoscale Science, Cornell University, Ithaca NY 14853

*rtc77@cornell.edu

**Abstract:**

The long-missing polarization-induced two-dimensional hole gas is finally observed in undoped gallium nitride quantum wells. Experimental results provide unambiguous proof that a 2D hole gas in GaN grown on AlN does not need acceptor doping, and can be formed entirely by the difference in the internal polarization fields across the semiconductor heterojunction. The measured 2D hole gas densities, about $4 \times 10^{13}$ cm$^{-2}$, are among the highest among all known semiconductors and remain unchanged down to cryogenic temperatures. Some of the lowest sheet resistances of all wide-bandgap semiconductors are seen. The observed results provide a new probe for studying the valence band structure and transport properties of wide-bandgap nitride interfaces, and simultaneously enable the missing component for gallium nitride-based p-channel transistors for energy-efficient electronics.

**7/23/2018**

The discovery of p-type impurity doping of the wide-bandgap semiconductor gallium nitride (GaN) around 1990 changed the field of semiconductor physics (*1*). It enabled the immediate realization of bright blue light emitting diodes and lasers, and started the solid-state lighting revolution, which today has transformed the lives of a large fraction of the population of the planet (*2*, *3*). To make energy-efficient visible lighting successful, it is necessary to inject both electrons and holes from supply layers in GaN into InGaN quantum wells, where they recombine and produce photons of desired wavelengths. This requires the complementary n-type doping of GaN too, which was fortunately available for several decades before the discovery of p-type impurity doping. While holes are generated by substitution of Ga atoms in the GaN crystal by Mg acceptor atoms, n-type doping is achieved by replacing Ga by Si or Ge donor atoms.

In the mid 1990s, high conductivity quantum-confined 2D electron gases were discovered at the heterointerface of AlGaN/GaN structures (*4*). Most remarkably, these 2D electron gases (2DEGs) did not require the presence of dopants. A few years after the observation, the reason for the formation of the 2DEG was tracked down to the existence of broken inversion symmetry in the GaN crystal, combined with the very high polarity of the metal-nitrogen bond in GaN and AlN (*5*, *6*). These properties lead to the existence of spontaneous and piezoelectric electronic polarization fields along the [0001] axis of the wurtzite nitride semiconductor crystal. The resulting polarization-induced 2D electron gas at Al(Ga)N/GaN heterojunctions has, in the last two decades, enabled high-voltage and ultra-high speed transistors that are being adopted in power electronics, and high-speed cellular communications in the RF and millimeter wave (*7*).

The p-type analog of the undoped polarization-induced 2D electron gas - the undoped 2D *hole gas*, however has remained elusive till today. Although low density 2D hole gases have been previously inferred in nitride heterojunctions in several reports (*8–15*), they have been either modulation Mg-doped heterostructures, or structures in which both electrons and holes are present. The missing dual piece of the *undoped* 2D hole gas has held back the widespread use of GaN for complementary logic electronics for digital applications till today, just like the absence of bulk p-doping had held back high efficiency photonic devices till the 90s. Significant advances in energy-efficient electronics can be enabled by GaN based high-voltage complementary low loss switches exploiting the large bandgap of the semiconductor, if a high conductivity undoped 2D hole gas can be found.

GaN and AlN demonstrate broken inversion symmetry along the [0001] axis or the c-direction of their wurtzite crystal structure, leading to the existence of spontaneous polarization $\boldsymbol{P_{sp}^{GaN}}$ and $\boldsymbol{P_{sp}^{AlN}}$ (*16*). This implies the existence of two distinct polarities: we consider metal-polar structures in this work. Because AlN has a smaller lattice constant than GaN, a thin epitaxial layer of AlN grown on top of a relaxed GaN layer is compressively strained, leading to a piezoelectric polarization $\boldsymbol{P_{pz}^{AlN}}$. The spontaneous and piezoelectric polarization fields add in the AlN layer, and the difference across the AlN/GaN heterojunction, $[(\boldsymbol{P_{sp}^{AlN}} + \boldsymbol{P_{pz}^{AlN}}) - \boldsymbol{P_{sp}^{GaN}}] \cdot \hat{\boldsymbol{n}} = \sigma_\pi$ is the net fixed polarization sheet charge density formed at the heterojunction. If the crystal is oriented in the metal-polar direction, this fixed polarization sheet charge is *positive* in sign. This polarization charge (and the resulting electric field), along with the electron potential energy barrier provided by the large conduction band offset $E_C^{AlN} - E_C^{GaN} = \Delta E_C$, induces the formation of the quantum-confined 2D electron gas at such a heterojunction. The densities that can be induced by the polar discontinuity are limited only by the polarization sheet charge $\sigma_\pi$, and far exceed those achieved by modulation doping or Mott criteria, and do not cause ionized impurity scattering. Such robust polarization induced 2DEGs in Al(Ga)N/GaN heterojunctions have been investigated for the last two decades and contributed to several applications such as ultrafast unipolar transistors and sensors (*7*, *17*).

If on the other hand, a thin layer of GaN is grown epitaxially on a relaxed AlN substrate, the GaN layer is under tensile strain. For the metal-polar orientation, the polarization difference $[(\boldsymbol{P_{sp}^{GaN}} + \boldsymbol{P_{pz}^{GaN}}) - \boldsymbol{P_{sp}^{AlN}}] \cdot \hat{\boldsymbol{n}} = \sigma_\pi$ is *negative* in sign. This negative immobile interface polarization charge should induce positively charged mobile carriers, or holes. The valence band offset of AlN and GaN, $E_V^{AlN} - E_V^{GaN} = \Delta E_V$, provides the necessary barrier for quantum-confining the holes to 2D. This is schematically shown in the energy band diagram shown in Figure 1(a), which is a self-consistent solution of a multiband quantum-mechanical k.p, and Poisson equations (*18*) for the GaN/AlN heterostructure. A mobile 2D hole gas of sheet density close to the fixed interface polarization charge $\sigma_\pi \sim 4 \times 10^{13}$ cm$^{-2}$ is expected to form at the heterojunction, depending on the thickness of the GaN layer. The holes are formed due to the field-ionization (or quantum tunneling) of electrons out of the valence band states into empty, localized surface states.

Figure 1(b) shows the layer structures that were grown for this study. A metal-polar AlN surface on a c-plane sapphire crystal was used as the substrate. An GaN/AlN layer was grown on it by molecular beam epitaxy (MBE). Figure 1(c) shows a zoomed in lattice image of the crystal heterointerface. A sharp heterojunction is observed, across which GaN and AlN are in the wurtzite crystal structure, and the GaN layer is coherently strained to the AlN layer. Further structural and chemical details of the heterojunction are shown in Figure 2. Figure 2 (a) shows a smooth surface morphology of the as-grown surface, with rms roughness less than 1 nm in a 10 um × 10 um scan area, and clearly resolved atomic steps. Figure 2(b) shows the X-ray diffraction spectrum of the heterojunction. The fringes and multiple peaks indicate a smooth few nm thick layer over the entire photon beam size of mms. This is further corroborated by the large width TEM images in the supplementary section Figure S1. Figure 2(c) is the reciprocal-space X-ray map, which proves that the GaN epitaxial layer is coherently strained to the underlying AlN layer, with an extracted biaxial compressive strain of 2.4 %. The strain state determines the net piezoelectric polarization charge in the heterostructure. Figures 1 and 2 thus collectively show that the heterostructure is structurally and chemically in a form that should exhibit the undoped polarization-induced 2D hole gas, and the transport studies discussed next indicate indeed this is the case.

Figure 3(a) shows the layer structure of two samples: Sample A is an undoped ~13 nm GaN layer on AlN. Sample B is identical to A, except the top 10 nm of GaN are doped with Mg to lock the surface potential. The doping screens any mobile carriers that may form at the buried heterojunction quantum well from the variations of the surface condition. For comparison with conventional chemical doping, a thick Mg-doped GaN (sample C), which is expected to have thermal ionization of holes (*19*), is the control sample in this study. Corner ohmic contacts were made to the three samples by using Indium dots in a van der Pauw geometry. Temperature-dependent Hall-effect transport properties of the three samples were measured from 300 K - 20 K.

Figures 3 (a, b, and c) show the measured data. The Hall-effect sign was observed to be *positive for all samples*, ensuring we are studying and comparing only holes in this study, and the interpretation is not clouded by parallel electron conduction. From Figure 3(a) it is seen that the resistivity of the Mg:GaN doped bulk control sample (Sample C) increases sharply with the lowering of temperature, from ~40 kOhm/sq at 300K to 2000 kOhm/sq at ~180 K. Fig 3(b) shows that this increase in resistivity in the control sample C is almost entirely due the decrease of the mobile hole density, which freezes out as $e^{-E_A/k_bT}$ with $E_A$~170 meV from ~$1.5 \times 10^{13}$ cm$^{-2}$ (bulk density of ~$4.1 \times 10^{17}$ cm$^{-3}$) at 300 K to ~$2 \times 10^{11}$ cm$^{-2}$ (bulk density of ~$5.5 \times 10^{15}$ cm$^{-3}$) at ~180

K. Thus, the thermally ionized holes freeze out, making the sample too resistive to measure below ~180 K. The hole mobility of sample C increases very nominally from ~10 cm$^2$/V.s at 300 K to ~15 cm$^2$/V.s at 180 K. On the other hand, a dramatically different behavior is seen for the undoped heterostructure sample A, and the same heterostructure with the Mg-doped cap layer Sample B. They are metallic, with the resistivity *decreasing* with temperature, showing the tell-tale signatures of a 2D hole gas.

Figure 3(a) shows that the resistivity of the undoped heterostructure Sample A decreases from ~11 kOhm/sq at 300 K to ~4 kOhm/sq at 20 K. The resistivity of the doped heterostructure Sample B decreases from ~8 kOhm/sq to ~2 kOhm/sq over the same temperature range. Figures 3(b) and 3(c) show that unlike the doped sample, the temperature dependences are flipped: the hole density in samples A and B are nearly independent of temperature, and all the change in the resistivity is due an increase in the hole mobility as the temperature is lowered. The hole sheet densities measured are nearly identical for the doped and undoped heterostructures in samples A and B. This would be impossible without the polarization charge at the interface because the integrated acceptor sheet density in sample B is only ~5 × 10$^{12}$ cm$^{-2}$, about an order of magnitude lower than the measured mobile hole gas density. This measurement constitutes the first unambiguous proof of the presence of a high-density polarization-induced 2D hole gas in undoped nitride heterojunctions. In these heterostructures, there simply are no other carriers such as parallel electrons channels or parallel 3D hole channels that can mask the direct and unambiguous measurement of the properties of the 2D hole gas. Although the Mg-doped cap layer offers very little mobile holes to the 2D hole gas, it can enable low-resistance tunneling p-type contacts to the 2D hole gas because of the high electric field it generates near the surface,.

Figure 3(c) shows that the mobility of the 2D hole gas in undoped Sample A and doped Sample B increases substantially, by ~6-9X from 300 K to 20 K. The mobility of the doped GaN/AlN heterostructure 2D hole gas increases from ~20 cm$^2$/V.s at 300 K to ~120 cm$^2$/V.s at 20 K, and from ~10 cm$^2$/V.s to ~90 cm$^2$/V.s for the undoped GaN/AlN heterostructure. The variation of the measured 2D hole gas mobility with temperature is expected to be strongly influenced by acoustic phonon scattering at all temperatures, in addition to the polar optical phonon scattering that dominates in most polar compound semiconductors. It is expected to depend sensitively on the effective mass of the valence bands near the Fermi level, which is affected by the biaxial compressive strain in the GaN layer at the heterojunction. At the lowest temperatures, when the phonon number is frozen out according the Bose-Einstein distribution, the interface roughness and impurity scattering should dominate. Though the hole mobilities do not saturate at ~20K, an extrapolation points to values in the range of ~100 - 200 cm$^2$/V.s. Since the 2D hole gas density survives to cryogenic temperatures, magnetotransport studies can directly access and probe the nature of the valence band of GaN in future studies.

The community has long thought that dislocations (as seen in the AFM image in Figure 2(a)) and other compensating centers whose formation are thermodynamically favored, such as Ga vacancies, preclude formation of 2D hole gases in GaN heterostructures. But experimental observations here show that the polarization discontinuity induces a 2D hole gas despite the presence of dislocations. Table S1 in the supplementary section also shows that the 2D hole gases are observed in multiple samples similar to Samples A and B with reproducible properties, constituting conclusive proof. As a further proof of the polarization-induced origin of the 2D hole gas, Figure S2 in the supplement shows that the electrical conductivity of the 2D hole gas varies with the thickness of the GaN layer, with a well-defined critical thickness. This is the exact dual of what is observed in the undoped polarization-induced 2D electron gas in Al(Ga)N/GaN heterostructures, and is a key stepping stone towards the realization of high-voltage p-channel transistors.

How do the observed polarization-induced 2D hole gases in the undoped and doped GaN/AlN heterostructures compare to those reported in nitride semiconductors, and in general to hole gases cutting across various semiconductor material systems? This is shown in Figures 4 (a) and (b). Figure 4 (a) shows that the 2D hole gas densities of $p_{2d} \sim 4 \times 10^{13}$ cm$^{-2}$ measured in this work in both the undoped and doped GaN/AlN heterostructures are close to the limit of the difference in polarization between AlN and compressively and coherently strained GaN. This is the dual of the 2D electron gas, where the corresponding limits are also seen in binary AlN/GaN heterojunctions (*20*). The hole densities are much higher than previously reported 2D hole gas densities in nitride semiconductors (*8–15*). In fact, the densities are among the highest among *all* semiconductor heterostructures, including SrTiO$_3$/LaAlO$_3$ (*21*), hydrogen-terminated diamond (*22–25*), strained Ge/SiGe (*26–29*), Si inversion channels (*30*), and GaSb/InGaAs (*31*) as shown in Figure 4 (b). The high 2D hole density in the nitride leads to some of the lowest sheet resistances, in spite of lower hole mobilities.

The 2D hole gas *mobilities* in the wide-bandgap nitrides are not on the high side because of the high valence band effective mass of both heavy and light holes in GaN due to its large bandgap. Ge/SiGe and GaSb/InGaAs heterojunctions show higher 2D hole gas mobilities due to smaller valence band effective masses, resulting from their small bandgaps. However, small bandgaps also mean limited capacity to handle high voltages, limiting them to low power applications. The large bandgap of the nitrides means that the high 2D hole gas densities can be modulated effectively with a gate, because the semiconductor intrinsically is capable of sustaining much larger electric fields. It is further conceivable that the hole mobility in the polarization-induced 2D hole gases in the nitride heterojunction could be improved by sharper interfaces, as the junctions studied here have binary semiconductors and have minimal alloy scattering, and scattering from dopants. But the most attractive way to improve the hole mobility may be by engineering the strain such that the valence bands are reordered such that a light effective mass is responsible for in-plane transport in the quantum well. Because the hydrogen present in the Metal-Organic Chemical Vapor Deposition (MOCVD) growth environment forms a complex with Mg, inhibiting its capacity to provide holes in GaN, buried p-type layers have been impossible by this popular growth technique because the $H_2$ does not diffuse through n-type layers. Though we have used plasma-MBE growth in this study, the fact that high-density holes are generated *without* Mg doping suggests that this technique can potentially be achieved also by MOCVD. Because of the fundamentally different origin of the 2D hole gas in the nitrides in the intrinsic polarization fields from broken inversion symmetry, this form of doping is expected to scale down to the individual unit cells, and not be affected by random dopant fluctuations. Future generations of small transistors can take advantage of, and someday depend on this unique scaling property of polarization-induced doping - now available in both the n-type and p-type recipes.

The discovery thus offers an attractive, clean, and technologically relevant platform to study the materials science and physics emerging in wide-bandgap and polar semiconductor heterostructures due to very large built-in electric fields. Strong effects of tunneling and Rashba-induced spin-orbit coupling are expected in these structures. The first unambiguous observation of the elusive polarization-induced 2D hole gas in undoped nitride semiconductor heterostructures thus completes a long search for its existence. Together with its dual, the polarization-induced 2D electron gas, wide-bandgap complementary logic electronics is now within reach.

*Sci. Technol.* **19**, L106–L109 (2004).

29. H. von Känel, D. Chrastina, B. Rössner, G. Isella, J. P. Hague, M. Bollani, High mobility SiGe heterostructures fabricated by low-energy plasma-enhanced chemical vapor deposition. *Microelectron. Eng.* **76**, 279–284 (2004).

30. M. Kaneko, I. Narita, S. Matsumoto, The Study on Hole Mobility in the Inversion Layer of P-Channel MOSFET. *IEEE Electron Device Lett.* **6**, 575–577 (1985).

31. S. Shin, Y. Park, H. Koo, Y. Song, J. Song, GaSb/InGaAs 2-dimensional hole gas grown on InP substrate for III-V CMOS applications. *Curr. Appl. Phys.* **17**, 1005–1008 (2017).
**Acknowledgments:** The authors thank Dr. Vladimir Protasenko, Dr. Henryk Turski, Dr. S.M. Islam and Ryan Page for help with the operation of the molecular beam epitaxy system, and Dr. Antonio Mei for assistance with the X-ray diffraction reciprocal space mapping and data analysis. **Funding:** This work was supported partly by Intel, Air Force Office of Scientific Research (Grant AFOSR FA9550-17-1-0048) monitored by Dr. Ken Goretta, and National Science Foundation (Grants 1710298 monitored by Dr. Tania Paskova and 1534303 monitored by Dr. John Schlueter). Characterizations and measurements were performed in part at Cornell NanoScale Facility, an NNCI member supported by NSF (Grant ECCS-1542081). The presented data made use of the Cornell Center for Materials Research Shared Facilities which are supported through the NSF MRSEC program (DMR-1719875). Z.C. is funded through PARADIM as part of the NSF Materials Innovation Platform program (DMR-1539918). **Author contributions:** R.C. grew and characterized the GaN/AlN epitaxial structures, with measurement assistance from S.J.B. S.J.B. performed the numerical simulations for the study. Z.C. performed scanning transmission electron microscopy (STEM) analysis of the epitaxially grown heterostructure, under the supervision of D.A.M. R.C. performed experimental data analysis, with help from D.J., S.J.B and H.X. R.C. and D.J. wrote the manuscript with inputs from all authors. **Competing interests:** The authors have applied for a patent on polarization-induced 2D hole gas for high-voltage p-channel transistors. **Data and materials availability:** All data is available in the main text or the supplementary materials.
*10*

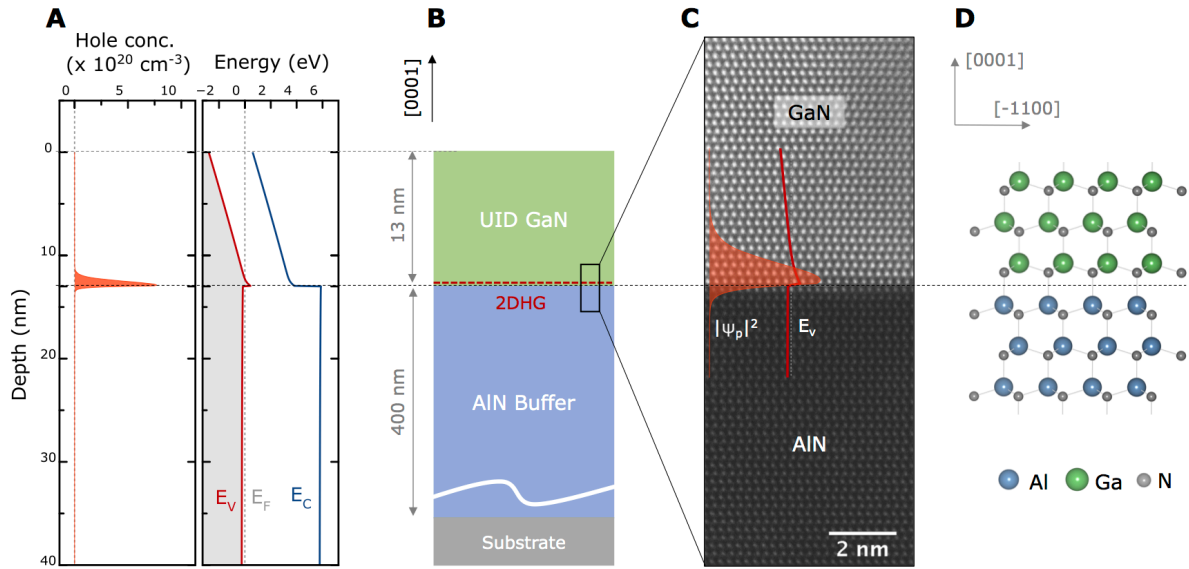

**Fig. 1.** Epitaxially grown GaN/AlN heterostructure. (**A**) Simulated energy-band diagram of a 13 nm undoped GaN epitaxial layer on AlN, showing the triangular quantum well formed in the valence band, and the high-density confined holes accumulated at the GaN/AlN interface. (**B**) Schematic of the epitaxially grown layer structure. (**C**) High resolution scanning transmission electron microscopy (STEM) image showing the wurtzite crystalline lattice, and the sharp GaN/AlN heterointerface. The valence band edge, and probability density of the holes from (**A**) are overlaid on the interface. (**D**) A schematic illustration of the Ga-polar GaN/AlN lattice interface and the expected 2DHG.

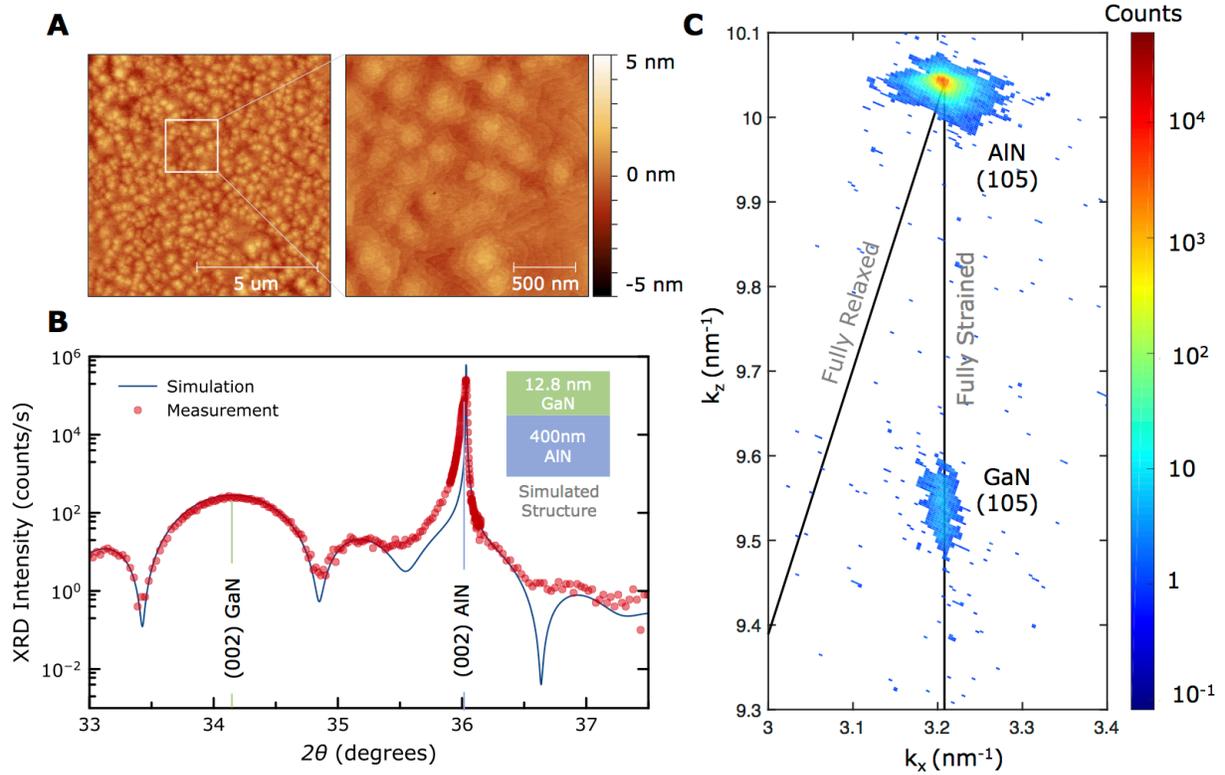

**Fig. 2.** Structural properties of the MBE-grown GaN/AlN heterostructures. (**A**) Atomic Force Microscopy (AFM) scans of the as-grown surface. The rms roughnesses are ~0.699 nm and 0.463 nm for the 10um and 2um scan respectively, and atomic steps are clearly seen. (**B**) X-ray diffraction (XRD) 2θ scan across the symmetric (002) reflection and the simulated data, confirming the targeted thicknesses and sharp interfaces. (**C**) Reciprocal space map (RSM) scan of the asymmetric (105) reflections of GaN and AlN shows the 13 nm GaN layer is fully strained to the AlN layer. Lines corresponding to fully strained and fully relaxed GaN are overlaid for reference.

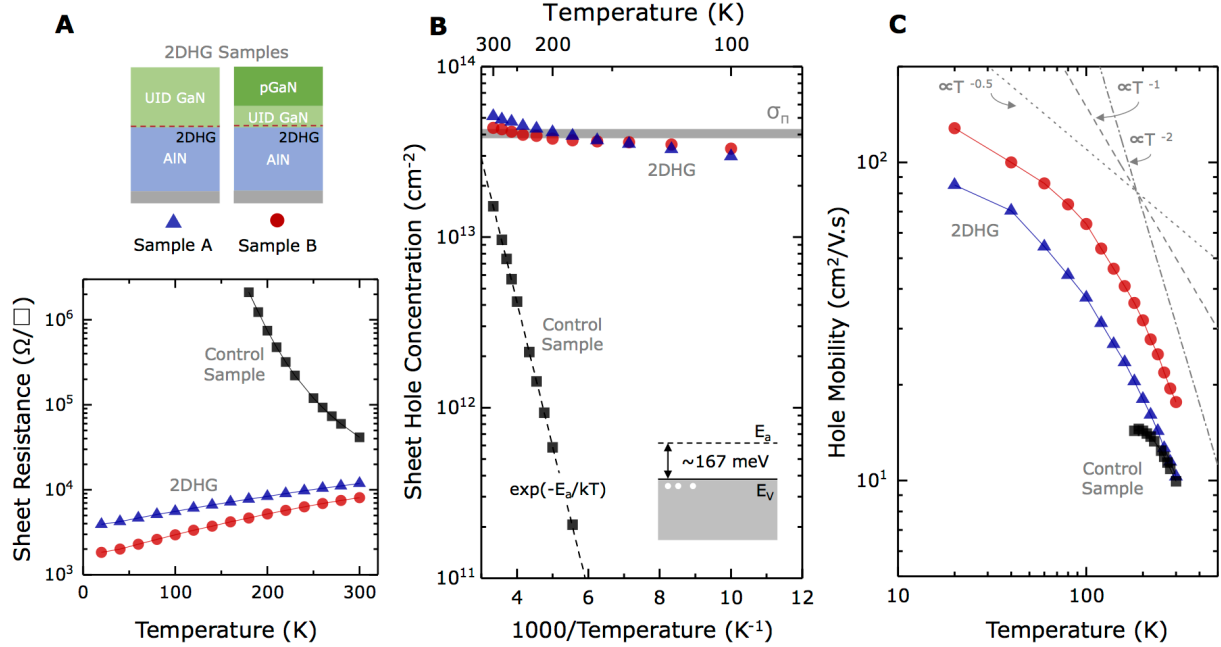

**Fig. 3.** Results of temperature-dependent Hall-effect measurements from 300K to 20K at 1T magnetic field of 2DHG samples A and B, along with a Mg-doped GaN control sample C. (**A**) The 2DHG samples A and B exhibit a metallic behavior of decreasing sheet resistance with decreasing temperature, whereas the control sample A is insulating in behavior, with a sharp increase in sheet resistance, and becoming too resistive at ~180K for measurement. (**B**) The measured mobile hole concentrations over a range of temperatures in samples A, B, and C. In the Mg-doped GaN (Sample C), holes freeze out below 180K. The density in the 2DHG of Samples A and B show almost no change in the hole concentration down to cryogenic temperatures. (**C**) The measured hole mobilities in Samples A, B, and C for the range of temperatures. While the control bulk doped sample shows a peak in mobility and then freezeout below 180K, the 2DHG in Samples A and B show significantly higher mobilities than C, with 2DHG mobilities reaching ~120 cm$^2$/V.s at 20K.

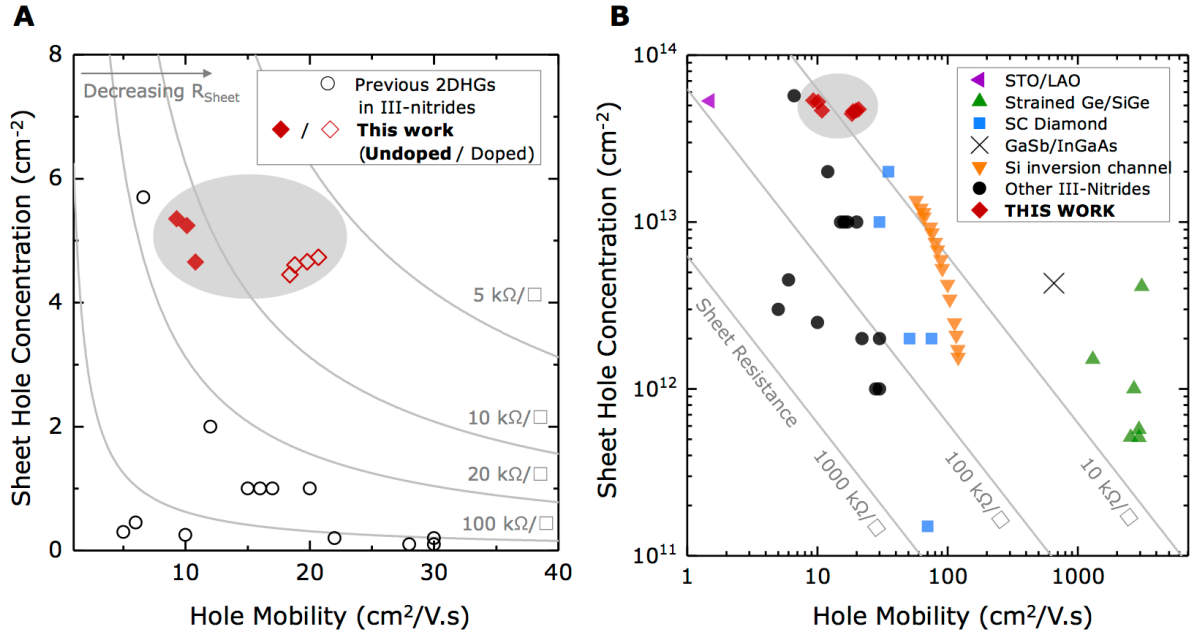

**Fig. 4.** Comparison of room temperature transport properties of 2D hole gases. (**A**) Comparison of the 2DHG mobility and sheet density reported in this work, with previously reported 2DHGs in nitride heterostructures (hollow symbols indicate Mg-typed doped). The doped as well as undoped structures reported in this work have much higher hole densities and decent mobilities, enabling record high p-type conductivity of ~10 kOhm/sq. (**B**) Comparing across various semiconductor material systems and heterostructures, the GaN/AlN 2DHG reported in this work has the highest room temperature hole density, and among the highest high conductivities for wide-bandgap semiconductors. The data for the doped/undoped 2DHG structures plotted here are shown in Table S1 of the supplementary materials.

**Supplementary Materials for** *A polarization-induced 2D hole gas in undoped gallium nitride quantum wells* **by Chaudhuri R.** *et al.*

**Materials and Methods**

The III-nitride heterostructures studied in this work were epitaxially grown in a Veeco Gen10 plasma-assisted molecular beam epitaxy (PA-MBE) system. Both the doped and undoped GaN/AlN structures were grown on a starting substrate of commercially available 1 micron thick semi-insulating Al-face [0001] AlN on c-plane sapphire templates from DOWA. 8mm × 8mm diced substrate pieces were ultrasonicated in acetone, methanol and isopropanol in succession before being mounted on a 3-inch lapped Si carrier wafer using molten Indium. The samples were then loaded into the MBE system, and outgassed at 200°C for 8 hours in a load-lock chamber, followed by at 500°C for 2 hours in a clean preparation chamber. They were then introduced into the MBE growth chamber and heated to the desired growth temperature. Effusion cells filled with ultra-high-purity sources were used for Ga, Al and Mg beams, whereas a RF plasma source with ultra-high-purity $N_2$ gas flowing through a mass-flow-controller and a purifier was used to provide the active N flux. The entire heterostructures reported in this work were grown at a $N_2$ RF power of 400W, resulting in a growth rate of ~560 nm/hr. A ~400 nm thick AlN buffer was first grown at a thermocouple temperature $T_{TC}$ = 780°C, with an effective beam-equivalent Al flux of ~$9 \times 10^{-7}$ Torr. Care was taken while growing the AlN buffer to reduce impurity contamination from the substrate surface. In order to maintain an abrupt heterointerface for the 2D hole gas, the excess Al on the growth surface was consumed by opening only the N shutter and monitoring the reflection high energy electron diffraction (RHEED) intensity until it saturated. The sample was then cooled to $T_{TC}$ = 730°C for the growth of the GaN layer. Approximately ~13 nm of unintentionally doped GaN was grown under an effective Ga flux of ~$1.0 \times 10^{-6}$ Torr. For the undoped structures, the Mg source was kept cold, and the shutter closed throughout, to avoid unintentional Mg doping of the GaN cap layers. On the other hand, for the doped GaN/AlN structures, during the last ~10 nm of GaN growth, the Mg source shutter was opened to incorporate Mg acceptor dopants. The Mg acceptor concentrations $N_A$ in the doped samples were verified to be between ~$5 \times 10^{18}$ cm$^{-3}$ to ~$1 \times 10^{19}$ cm$^{-3}$ in different samples, as calibrated by secondary ion mass spectrometry (SIMS) measurements performed on a separate Mg doping calibration sample grown under the same epitaxial conditions.

Atomic force microscopy (AFM) scans performed in a Bruker ICON Dimension system after MBE growth showed a smooth surface with sub-nm rms roughness over both a large area 10 um × 10 um scan and smaller area 2 um × 2 um scans. X-ray diffraction measurements were performed in a Panalytical XRD system using the Cu-Kα line source. A 2theta scan along the (002) symmetric peak of the AlN/GaN structure showed the AlN and GaN reflection peaks and confirmed the thicknesses of the layers by comparing and fitting to a simulated model diffraction spectrum. The reciprocal space map (RSM) around the (105) asymmetric reflection of the epitaxial heterostructure was examined to extract the following in-plane and out-of-plane lattice constants: $a_{AlN}$ = 0.311 nm, $a_{GaN}$ = 0.311 nm and $c_{AlN}$ = 0.4985 nm, $c_{GaN}$ = 0.5241 nm. Comparing to the unstrained lattice constant, the GaN layer is under ~2.41% compressive strain. Scanning transmission electron microscopy (STEM) was performed on a probe aberration corrected FEI Titan Themis operating at 300 keV. Thin cross-sectional samples along [110] and [$\bar{1}$10] crystal orientations were prepared using focused ion beam (FIB) and imaged using the annular dark field (ADF) mode of the STEM. The wide area scans shown in **Figure S1** confirmed the abrupt interface between the GaN and AlN.

Hall-effect measurements at 300 K and 77 K were first performed on all the grown samples using a Nanometric Hall-Effect System. The transport data of a selection of samples are tabulated in **Table S1**. These samples were grown over different growth days and illustrate the high repeatability of the 2D hole gas. These data are included in the benchmark plot in Figure 4. One of the doped heterostructure, and one undoped heterostructure (samples A2 and B4 in Table S1) were further characterized by measuring the temperature dependent Hall effect from 300 K to 20 K, at 1T magnetic field in a Lakeshore closed-cycle cryogenic stage. The data is presented in the main text Figure 3.

**Figure S2** shows the measured variation of the 2D hole gas conductivity with changing GaN layer thickness for an undoped 2DHG heterostructure. **Figure S2 (a)** shows the expected change with thickness in the energy band diagram of the valence band, and the mobile hole concentration with depth, as simulated using a self-consistent multiband ***k.p*** Schrodinger-Poisson solver *(18)*. A typical bare-GaN surface conduction band edge barrier height of 0.7 eV was used for the simulations. By combining the simulations with experimentally measured hole mobilities, we can obtain the 2D hole gas conductivity expected as a function of GaN thickness and temperature. The solid lines in **Figure S2 (b)** show these expected results, using hole mobilities of 10/22 cm$^2$/V.s at 300 K/77 K respectively for calculating the hole sheet conductivity. The solid lines indicate a critical thickness of ~5 nm below which the 2DHG is depleted from the surface potential, a sharp rise from ~5 - 20 nm, beyond which the hole density saturates to the interface polarization sheet density. The measured 2D hole gas conductivity should follow a similar trend. To test this, a thick 30 nm undoped GaN on AlN sample was grown. It was successively dry-etched using low power, a low damage RIE/ICP etch process to the desired thicknesses. The sheet resistances were measured by Hall-effect after every etch step at both 300 K and 77 K. The measured conductivity is plotted in **Figure S2 (b)** alongside the solid lines predicted from the polarization discontinuity and a fixed surface barrier height. The qualitative agreement to the simulated model is a further proof that the 2D hole gas is indeed polarization-induced.

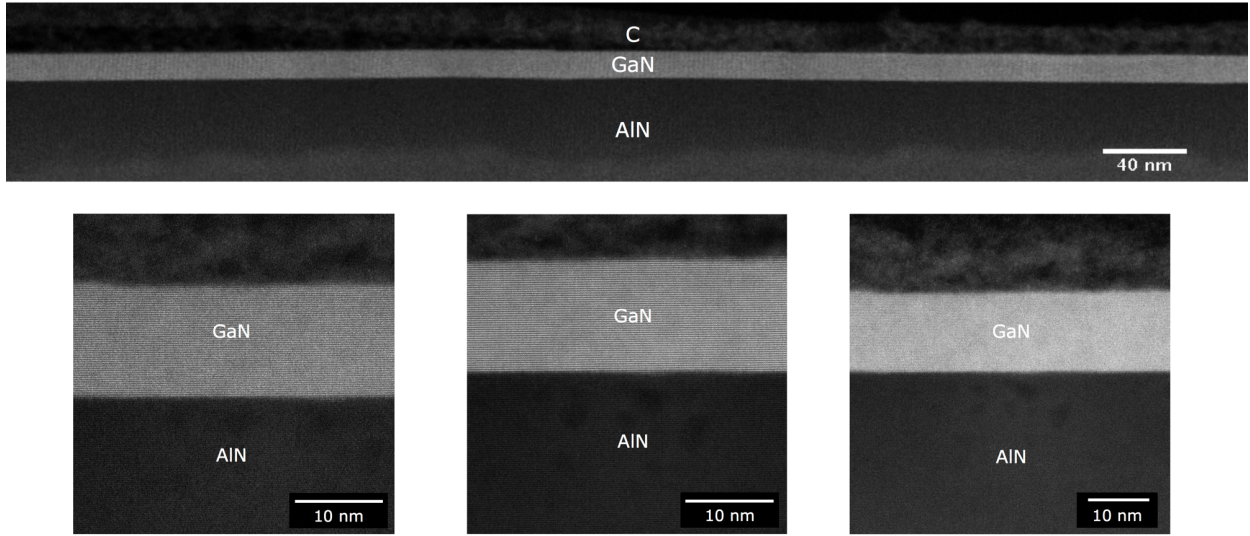

**Fig. S1.**

STEM annular dark field (ADF) images of the cross-section of the GaN/AlN heterostructure along [$\bar{1}$10] zone-axis. The wide-area image and the zoomed-in regions clearly show the coherently sharp interface between the GaN and AlN (~1-2 ML) is maintained over large areas of the wafer, which is essential for a high mobility and high uniformity of the polarization-induced 2D hole gas over the entire wafer.

*18*

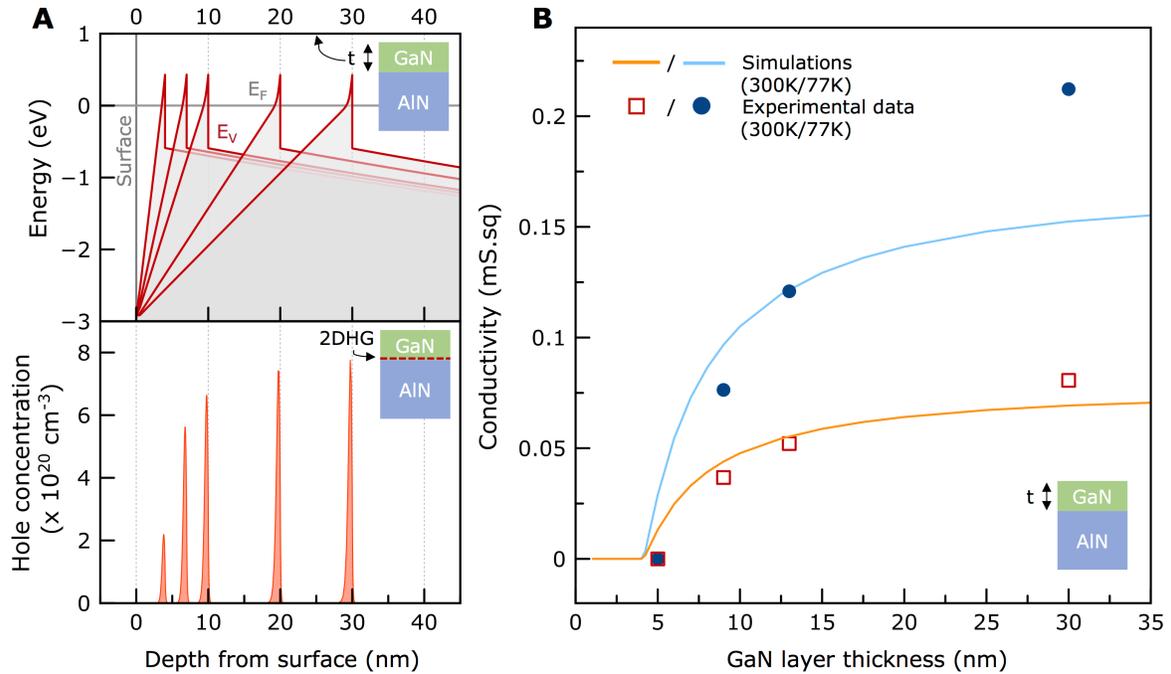

**Fig. S2.**

Dependence of the properties of the polarization-induced 2D hole gas on the thickness of the undoped strained GaN layer grown on AlN. (**A**) shows the dependence of the valence band edge and the spatial hole density distribution as a function of the GaN thickness. The triangular quantum well at the heterojunction is clearly visible, with the large valence band offset to confine the 2D holes to a width of ~1 nm in the vertical direction. The 2D hole gas density increases and saturates for GaN layer thicknesses above ~30 nm. (**B**) shows the numerical simulations for the variation of the 2D hole gas conductivity with the thickness of the GaN cap layer. The experimentally measured values are in good agreement to the model. To obtain the experimental data, the thickness of the GaN layer is changed by successively etching the GaN layer down from its original thickness of 30 nm. The conductivity is measured using Hall-effect measurements after each etch step. There clearly exists a critical minimum thickness of GaN for the existence of mobile holes at the GaN/AlN interface, and is a proof of the polarization-induced nature of the 2D hole gas.

| Sample | $N_{Mg}$ in GaN (cm$^{-3}$) | Temp (K) | $\mu_{Hall}$ (cm$^2$/V.s) | σ (cm$^{-2}$) | $R_S$ (Ω/□) |
|---|---|---|---|---|---|
| A1 | undoped | 300 | 9.28 | 5.35E+13 | 1.26E+04 |
|  |  | 77 | 40.3 | 3.14E+13 | 4927 |
| A2* | undoped | 300 | 10.1 | 5.24E+13 | 1.18E+04 |
|  |  | 77 | 48 | 2.74E+13 | 4751 |
| A3 | undoped | 300 | 10.8 | 4.66E+13 | 1.24E+04 |
|  |  | 77 | 26.3 | 5.03E+13 | 4715 |
| B1 | 1x10$^{19}$ | 300 | 19.8 | 4.66E+13 | 6766 |
|  |  | 77 | 78.8 | 3.81E+13 | 2081 |
| B2 | 1x10$^{19}$ | 300 | 18.8 | 4.61E+13 | 7221 |
|  |  | 77 | 68.3 | 3.80E+13 | 2406 |
| B3 | 1x10$^{19}$ | 300 | 20.7 | 4.73E+13 | 6364 |
|  |  | 77 | 71.2 | 3.93E+13 | 2232 |
| B4* | 5x10$^{18}$ | 300 | 18.8 | 4.37E+13 | 7595 |
|  |  | 77 | 102 | 3.72E+13 | 1647 |

**Table S1.**

Hall-effect measurement data for several MBE grown 13 nm GaN on AlN heterostructures showing the reproducibility of the 2D hole gas properties in undoped and doped structures. Samples A1-A3 are undoped heterostructures, whereas samples B1-B3 have ~10nm thick Mg-doped p-type GaN caps on the top of a ~3 nm undoped GaN layer grown on AlN. Samples A2 and B4 were also measured using temperature dependent Hall effect, and are presented in the main text in Figure 3. The data clearly shows the presence of a highly repeatable 2D hole gas, both in the doped and undoped heterostructures.